# Privacy Concerns and ChatGPT: Exploring Online Discourse through the Lens of Information Practice on Reddit

S M Mehedi Zaman, Saubhagya Joshi, Yiyi Wu

Rutgers University School of Communication and Information, USA

## Abstract

**Introduction.** As millions of people use ChatGPT for tasks such as education, writing assistance, and health advice, concerns have grown about how personal prompts and data are stored and used. This study explores how Reddit users collectively negotiate and respond to these privacy concerns.

**Method.** Posts were collected from three major subreddits — r/Chatgpt, r/privacy, and r/OpenAI — between November 2022 and May 2025. An iterative keyword search followed by manual screening resulted in a final dataset of 426 posts and 1,900 comments.

**Analysis.** Using information practice as the theoretical lens, we conducted a qualitative thematic analysis to identify collective practices of risk negotiation, validated with BERTopic topic modeling to ensure thematic saturation.

**Results.** Findings revealed risk signaling, norm-setting, and resignation as dominant discourses, and collective troubleshooting and advocacy for privacy-preserving alternatives as key adaptive practices.

**Conclusion(s).** Reddit functions as a site of collective sense-making where users surface risks, establish informal norms, and share strategies for mitigating privacy threats, offering insights for AI design and privacy literacy initiatives.

## Keywords

ChatGPT, Information Practice, Privacy, Reddit.

## Introduction

As millions of people use ChatGPT for everyday use, such as personal assistance (Ray, 2023), educational and writing assistance (Kasneci et al., 2023), seeking information (Dwivedi et al.,2023), entertainment through storytelling (Bianchi, 2024), and many more, some users have grown concerned about the privacy policy of this tool and how personal data and prompts given to the AI are handled by OpenAI (Colville & Ostern, 2024). This concern is valid as users do not know what is happening at the backend (how the tool is processing personal information, which data it is storing, with whom data is shared, etc.) when we agree to the terms and conditions of this tool (Amarikwa, 2024). Such opacity motivates negative public opinions about generative AI, skepticism around personal data storage and sharing with third-party companies for advertising, training of AI models with personal information, and the overall governance of OpenAI, all of which can be observed online on platforms like Facebook, Reddit, and X (Qi et al., 2024).

This research examines public opinions on Reddit through some of the most active subreddits — selected based on number of users and engagement — where people express concerns and exchange advice about privacy. On reddit, users generally join a subreddit relevant to their interests and post questions, opinions and concerns as posts, to which other users answer questions, share their opinions and address the original poster's (OP) concerns through comments, upvotes, and downvotes (Anderson, 2015). Hence, Reddit can be seen not only as a space for individual information seeking but also as a collective site where norms are shaped and information practices are negotiated (Record et al., 2018).

Several studies have examined negotiated privacy in online environments (Jarecki et al., 2002; Seamons et al., 2002; Sarvepalli, 2020). However, with the rapid rise of generative AI, it has become increasingly important to understand and analyze public discourse around privacy concerns related to these technologies. Hence, this research employs the lens of **information practice** (Savolainen, 2007) — emphasizing the socially situated, discursive, and negotiated nature of information activities — to analyze Reddit discourse about ChatGPT privacy. To this end, the following are the Research questions (RQs) we are seeking answers for:

1. What are the shared concerns and discourses that shape Reddit users' collective understanding of privacy risks with ChatGPT?
2. How do Reddit users collaboratively negotiate and adapt their information practices in response to these concerns?

## Literature Review

The rise of generative AI systems such as ChatGPT has prompted increasing scholarly attention to their applications, challenges, and ethical implications. Ray (2023) offers a comprehensive review of ChatGPT, highlighting its diverse uses but also underscoring the risks of bias, data misuse, and limited transparency in how user inputs are handled. Similarly,

Kasneci et al. (2023) explore both the opportunities and challenges of deploying large language models in educational contexts, arguing that while such tools can enhance accessibility and creativity, they introduce new privacy and fairness concerns that must be addressed at both technical and policy levels. Dwivedi et al. (2023) further emphasize that the implications of generative AI are not merely technical but extend to research ethics, policymaking, and public trust, noting that the question of "who owns the outputs and the data" remains unresolved.

Beyond technical and policy discussions, scholars have begun to investigate public perceptions of AI systems. Qi et al. (2024) analyze social media data to decode the balance between excitement and concern in the "post-ChatGPT era," finding that privacy and security issues are among the most frequently cited worries. This aligns with Colville and Ostern's (2024) argument that trust and distrust in generative AI are strongly mediated by users' AI literacy and their understanding of how data flows through these systems. Bianchi (2024) similarly stresses the need for pedagogical efforts that equip users with critical thinking skills to engage with AI tools responsibly. Together, these works suggest that addressing privacy concerns requires not only better technical safeguards but also interventions aimed at raising awareness and supporting informed decision-making.

In parallel, research on online communities provides insight into how users negotiate privacy risks collectively. Reddit has been studied as a site of health information seeking (Record et al., 2018) and general knowledge exchange, with its pseudo-anonymous nature encouraging candid discussion of sensitive topics (Anderson, 2015). Prior studies of "negotiated privacy" (e.g., Marwick & boyd, 2014) show that users engage in dynamic trade-offs between disclosure and protection, often developing shared norms for what information is safe to share. However, most of this literature predates the widespread adoption of generative AI and does not fully account for the unique challenges posed by AI systems that continuously learn from user data.

Taken together, existing research establishes the importance of understanding privacy concerns in AI use and the role of online communities in shaping information behavior. Yet, there remains a gap in examining how these concerns are collectively constructed, debated, and acted upon in the context of generative AI. This study addresses that gap by using an information practice framework to analyze Reddit discourse on ChatGPT privacy, focusing on how users collectively signal risks, set norms, and co-construct mitigation strategies.

## Methodology

This study adopts **Information Practice** as the central theoretical framework to analyze the discourse happening on Reddit regarding our area of interest. Information practice provides a way to focus on how users collectively make sense of privacy risks, negotiate shared norms, and co-construct strategies to mitigate concerns, rather than viewing privacy merely as an individual information-need problem.

The choice of Reddit as the data source is threefold – it is pseudo-anonymous (users feel more comfortable sharing their opinions on sensitive topics) (Anderson, 2015), data scraping is easier compared to platforms like Facebook or X, and it shows a high level of engagement. Data scraping was approved by our Institutional Review Board (IRB) with the condition that no direct user quotes be included in dissemination. Data collection proceeded in several steps:

1. Searched Reddit using the phrase *"privacy concerns ChatGPT OpenAI."*
2. Identified the top 10 subreddits from the results and noted the number of users and active participation levels.
3. Applied a threshold of at least 100 active users at any given time and retained only those meeting this criterion.
4. Finalized three subreddits — **r/Chatgpt**, **r/privacy**, and **r/OpenAI** — as the focus of the study.

An initial search using terms such as *privacy*, *chatgpt*, *concerns*, *trust*, and *security* yielded approximately 1,000 posts. Each post was manually reviewed, resulting in a filtered set of 150 posts (with approximately 1,000 comments) from November 2022 to May 2025. Based on this initial exploration, additional keywords such as *distrust*, *sensitive*, and *private* were used to broaden the search, yielding an additional 1,300 posts. After a second round of manual review, a final corpus of **426 posts** and approximately **1,900 comments** were curated for analysis.

We conducted a qualitative thematic analysis focusing on how users collectively expressed, debated, and acted upon privacy concerns.

- **Open Coding:** Read all posts/comments and coded expressions of concern, advice, and mitigation strategies.
- **Theme Development:** Grouped codes into collective practices such as *risk signaling*, *troubleshooting*, and *advocacy for privacy-preserving alternatives.*
- **Cross-Check:** Used BERTopic to validate that these themes reflected dominant discussion clusters.

## Findings

To confirm the reliability of our qualitative analysis, we utilized BERTopic to computationally map the underlying structure of the dataset. This automated topic modeling revealed distinct clusters of discussions such as specific groupings of terms related to data retention, security settings, and local hosting—that closely mirrored the themes we identified through manual coding. The strong alignment between the machine learning output and our human interpretation confirms that we achieved thematic saturation. It provides evidence that the categories described below are not merely isolated anecdotes, but accurately represent the most prevalent and significant patterns within the broader corpus of our posts and comments.

Our analysis revealed several recurring ways users expressed and framed their privacy concerns:

- Risk Signaling & Awareness-Raising: Users frequently engaged in risk signaling and awareness-raising, posting explicit warnings about the potential for data retention and backend access. Messages such as "Be careful what you type — it can be stored!" were common triggers for extended comment chains where others shared similar realizations or fears. These threads often served as a form of crowd-amplified alert system, helping users collectively surface risks that might otherwise go unnoticed.
- Norm Setting & Community Guidance: Over time, individual warnings evolved into informal community rules of caution. Users reminded one another to treat ChatGPT like a public platform, writing things like "Assume everything is saved somewhere." These reminders not only validated concerns but also helped establish a shared baseline of expectation for how to use the tool responsibly.
- Resignation & Pragmatic Acceptance: Some discussions reflected resignation and pragmatic acceptance of privacy risks. For a subset of users, the loss of control over data was reframed as an inevitable aspect of using a popular AI service — a kind of "cost of convenience." Rather than withdrawing from the tool, these users advocated a practical mindset, signaling that constant worry was unproductive and that participation came with unavoidable trade-offs.

Together, these themes show not just what concerns users have (e.g., data storage, backend opacity) but how they are collectively articulated and reinforced through shared discourse.

Beyond expressing concern, users actively engaged in collective problem-solving to adapt their behaviors:

- Collective Troubleshooting & Knowledge-Building: Reddit threads functioned as peer-to-peer help spaces where users pooled knowledge, compared settings, and sought reassurance about their chosen strategies. One person might share a step-by-step mitigation guide, such as disabling chat history, and others would respond with confirmation, clarifications, or alternative suggestions. Through these exchanges, a shared understanding of what "works" gradually emerged.
- Advocacy for Privacy-Preserving Alternatives: Technically skilled participants frequently suggested running local or open-source large language models (LLMs) as a way to regain control over personal data. These recommendations often shifted the tone of discussion from fear to empowerment, providing actionable solutions rather than merely voicing anxiety. In several cases, such suggestions inspired others to experiment with self-hosted models or change their own privacy settings.

These findings demonstrate how Reddit users collectively negotiate what counts as safe practice, build shared troubleshooting knowledge, and adapt their information practices in response to perceived privacy risks.

The findings also highlight Reddit as more than a space for individual question-asking — it operates as a site of collective privacy practice where users surface risks, negotiate norms, and develop shared strategies for navigating uncertainty around ChatGPT. By framing the results through the lens of information practice, we see that privacy concerns are not merely internal triggers but socially situated activities that unfold through discourse, peer validation, and interaction with platform affordances. The presence of risk signaling shows how users actively take responsibility for alerting others, transforming personal anxiety into community-level awareness. Norm setting demonstrates that users co-construct rules of engagement that govern how much risk is acceptable and what precautions should be taken, essentially forming an informal privacy literacy curriculum for others in the subreddit. At the same time, the discourse of resignation signals a tension between ideals of privacy and practical realities: some users are willing to accept data risks for the sake of convenience, suggesting that privacy interventions must be designed with usability and habit in mind to be effective.

The collective troubleshooting and advocacy for alternatives observed here underscore Reddit's role as a distributed knowledge network where users collaboratively experiment with mitigation strategies and share actionable solutions. These practices reflect an ongoing negotiation of power between users and technology providers: while platforms like OpenAI set the formal terms of use, users create their own counter-practices to regain agency over their data. Taken together, these patterns illustrate how online communities can act as early sites for detecting, framing, and responding to emerging privacy issues with generative AI. They also suggest that studying these interactions can provide valuable insight for designers, policymakers, and educators seeking to support informed, empowered, and contextually appropriate privacy practices.

## Conclusion

This research demonstrates that Reddit serves as more than a simple Q&A forum; it is a site of negotiation where individuals collectively navigate opaque technologies and build a shared understanding of risk. By viewing these interactions through the lens of information practice, we see that users' privacy anxieties are not merely gaps in their knowledge, but reflections of deeper structural and emotional tensions regarding trust, autonomy, and their relationship with Generative AI.

Consequently, these findings offer concrete directions for policymakers and AI developers to move beyond generic compliance and toward truly **"better" systems**.

- **What to Avoid:** Developers should avoid the "black box" design that currently fuels user paranoia—specifically, vague data retention policies and complex opt-out mechanisms that force users to rely on "folk theories" to feel safe.
- **What to Build:** Instead, a better system would prioritize granular control and transparency. For example, rather than a binary choice between using the tool and keeping data private, systems should offer visible toggles for data retention on a per-conversation basis or clear visual indicators of when data is being used for model training versus when it is ephemeral.

- **Policy Implication:** Policymakers should push for standards that do not force users into a resignation mindset, where they must trade privacy for utility. Instead, regulations should encourage hybrid architecture similar to the local LLMs advocated for by Reddit users—that allow sensitive processing to happen on the user's device while still accessing cloud capabilities for broader tasks.

Furthermore, these insights can empower activists and privacy organizations to better articulate the specific risks users face. Rather than abstract warnings, they can focus educational efforts on practical privacy literacy, helping users understand the actual mechanics of data storage versus the imagined risks. Finally, this research opens new avenues for scholars to examine how online discourse shapes the adoption of emerging technologies, with a particular focus on how community-driven risk signaling can compensate for the failures of official corporate communication.